\documentclass{AUJarticle}
\usepackage[cmex10]{amsmath}
\usepackage[utf8x]{inputenc}
\usepackage[nocompress]{cite}
\usepackage{graphicx, multirow, booktabs, color, listings, subcaption}
\usepackage{balance}

\usepackage{url}

\graphicspath{{figures/}}

\begin{document}

\title{Software-based security approach for networked embedded devices}

\addauthor{José Ferreira, Alan Oliveira, André Souto, José Cecílio}
{LASIGE, Departamento de Informática, Faculdade de Ciências da Universidade Lisboa,
Campo Grande 016, 1749-016 Lisboa}
  {fc53311@alunos.ciencias.ulisboa.pt,\{aodsa,ansouto,jmcecilio\}@ciencias.ulisboa.pt}


\issuev{1}
\issuen{1}
\issued{March 2023}

\shortauthor{J. Ferreira et. al}
\shorttitle{Software-based security approach for networked embedded devices}

\thispagestyle{plain}

\maketitle

\begin{abstract}
As the Internet of Things (IoT) continues to expand, data security has become increasingly important for ensuring privacy and safety, especially given the sensitive and, sometimes, critical nature of the data handled by IoT devices. 
There exist hardware-based trusted execution environments used to protect data, but they are not compatible with low-cost devices that lack hardware-assisted security features.
The research in this paper presents software-based protection and encryption mechanisms explicitly designed for embedded devices.
The proposed architecture is designed to work with low-cost, low-end devices without requiring the usual changes on the underlying hardware.
It protects against hardware attacks and supports runtime updates, enabling devices to write data in protected memory. 
The proposed solution is an alternative data security approach for low-cost IoT devices without compromising performance or functionality. Our work underscores the importance of developing secure and cost-effective solutions for protecting data in the context of IoT.

\end{abstract}

\section{Introduction}
\label{sec:intro}
As a result of the Internet of Things (IoT) popularization, millions of embedded devices are being deployed and connected to the worldwide network~\cite{verifyandrevive3427253}. 
These devices allow IoT to extend the boundaries of the Internet. At the same time, they connect digital processes to the physical world~\cite{microVisor8821489}. Although the resulting systems create added-value services for the respective applications, they inherently also bring vulnerabilities~\cite{pistis277124}. The threats must be mitigated since these systems might collect and process critical and sensitive data.

Among the techniques proposed to protect data are the Trusted Execution Environments (TEEs). TEEs are designed to provide mechanisms to protect applications (code and critical data), ensuring confidentiality and integrity~\cite{pistis277124}. However, most existing TEEs proposals rely on hardware~\cite{pistis277124}, such as Trusted Platform Modules (TPMs)~\cite{tpm14}, Intel SGX~\cite{intelsgx086} and ARM TrustZone~\cite{trustzone7809736}. Due to cost and size constraints, those hardware features mainly exist on high-end platforms and are unavailable on cost-effective and low-end embedded devices~\cite{pistis277124}. 

Alternatively, software-based approaches are being proposed, such as PISTIS~\cite{pistis277124}, Security MicroVisor ($S\mu V$)~\cite{microVisor8821489}, SofTEE~\cite{softee9131703}, and Virtual Ghost~\cite{virtualghost2541986}. These solutions have some advantages when compared with hardware-assisted ones. One advantage is related to update costs. It is easier and cheaper to update a software-based TEE. Another advantage is hardware portability, given that those TEEs do not require specific hardware features (such as ARM TrustZone or Intel SGX)~\cite{softee9131703}. Despite these advantages, most of those software-based secure architectures do not consider hardware-based attacks in their scheme, i.e., they do not consider scenarios in which the attacker has access to the device where the application is running and can change its code.

In the literature, it is possible to find solutions hybrid architectures, i.e., architectures that combine hardware and software modules to achieve the best of both solutions~\cite{vrased236230, smart3536, trustlite2592824, tytan7167218}. Although those solutions offer strong security guarantees for applications running on low-end devices, to be implemented in those devices, they require hardware modification~\cite{pistis277124}. Hardware modification is impracticable in real-world scenarios, considering that every device needs the addition of customized hardware~\cite{pistis277124}, and it is difficult for legacy devices to take advantage of them~\cite{softee9131703}. Moreover, the hybrid solution approaches usually consider that hardware-based attacks are out of scope.

Considering the lack of hardware security features in low-cost embedded systems and the benefits offered by software-based TEEs,
this work aims to develop a software-based security approach for networked embedded devices (SbS4NED) that provides a set of lightweight mechanisms to protect software and data integrity (continuously verifying the integrity of memory) and offer correction in case of unexpected changes. The application code will also be protected using encryption. This way, it becomes tempered resistant and offers more reliability to the verification process. 
Moreover, it will be supported by lightweight cryptography algorithms presented at the National Institute of Standards and Technology (NIST) competition~\cite{nist}. In particular, Xoodyak~\cite{xoodyak}, one of NIST’s finalists, will be used to encrypt data.

\section{Thread Model and Assumptions}
Defining the assumptions about the attacker's capabilities and goals, the system's components and interactions, and the security goals that need to be achieved are essential. In this work, we consider adversaries with the following capabilities:
\begin{itemize}
    \item The adversaries have access to the device. They may modify the application code running on the device to read or change the data the application handles.
    
    \item The adversaries can sniff the network, modify messages exchanged between devices, and perform man-in-the-middle attacks.

    \item Software-based adversaries may be present on the device where the architecture will be deployed. Their goal may be to change the data available in memory and consequently control the entities that rely on data accuracy. 
\end{itemize}

We assume that the SbS4NED is correctly installed on the devices by a trusted party. We also assume it is bug-free, encrypted, and working as expected. Therefore, the adversary can not surpass the code, and the verification process carried out by its components. The final assumption is that each device has mechanisms to compute the encryption key used to protect the local files where SbS4NED keys are stored.

\section{Software-based security Architecture} 

As mentioned in the Introduction, this work aims to build a software-based tempered-resistant solution that protects the software and data in networked
embedded devices. Driven by this goal, we design the SbS4NED proposed architecture.

\begin{figure}
    \centering
    \includegraphics[width=0.45\textwidth]{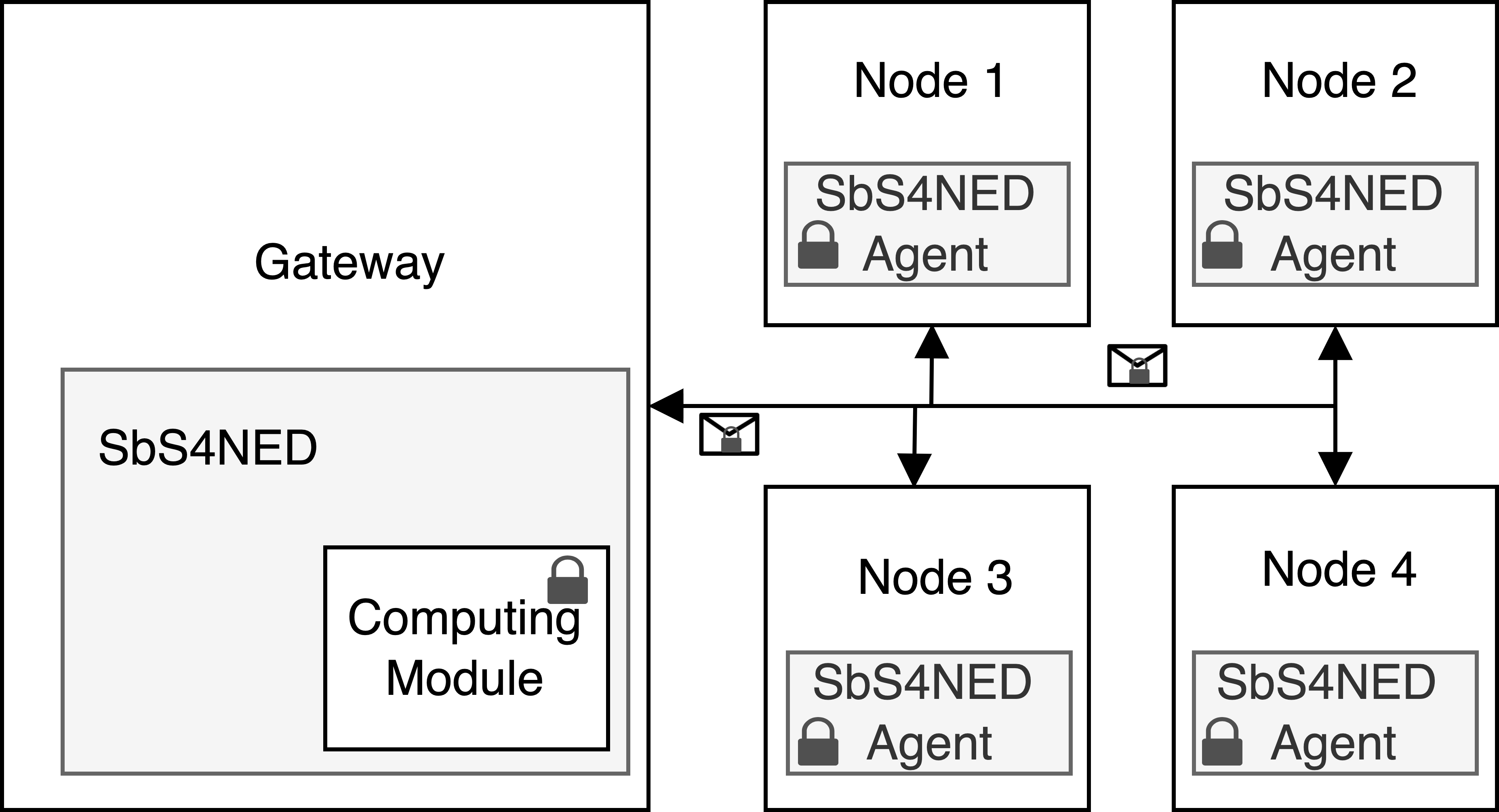}
    \caption{SbS4NED architecture.}
    \label{fig:prototype}
\end{figure}

Figure~\ref{fig:prototype} shows a high-level description of the proposed architecture, where the SbS4NED Computing Module (SbS4NED\_CM) runs inside the Gateway. It is responsible for monitoring applications running on the nodes connected to the Gateway. Each node will have an agent (SbS4NED\_Agent) generating the signature of the application code running on the node and sending it to the SbS4NED\_CM for code integrity check. The SbS4NED\_CM and the node code's application are encrypted to increase security and to offer more protection to SbS4NED\_CM internal processes. Moreover, the messages exchanged between SbS4NED\_CM and its agent are also encrypted. Next, we describe the SbS4NED components: 

\begin{itemize}
  \item \texttt{App Manager} -- It interacts with the applications deployed in the node and aims to perform application updates and send and receive data from the nodes. 
  
  \item \texttt{Key Manager} -- This component is responsible for managing ({\it i.e.}, generating and renewing) the keys used internally by SbS4NED\_CM and for external communication (with a SbS4NED\_Agent running on the node). It uses Diffie-Hellman (DH) key-exchange protocol for external communication to generate or renew the key.
  
  \item \texttt{Crypto} -- Provides cryptography services inside the SbS4NED\_CM and the SbS4NED\_Agent. It can encrypt, decrypt, and compute the message authentication code (MAC). In the SbS4NED\_CM side, the \texttt{App Manager} can also use this component to encrypt the app-compiled code before sending it to the node. This way, secure code update is ensured.
  
  \item \texttt{Integrity Checker} -- Designed for memory integrity checking. It writes the data from App Manager in the memory and holds the (randomized) position where it is written. The Integrity Checker is also responsible for remotely checking the integrity of the nodes' code.
  
  \item \texttt{Logger} -- It is responsible for keeping the log files updated regarding the memory integrity state, which app the data came from, which nodes are connected, and any network activity that must be logged to easily detect if an attacker is trying to join the network or injecting any data on the network.

  \item \texttt{App Thread} -- It is used for executing the application code developed by the user. It offers an API to interact with the node underlayer software and hardware layer. All the interaction must be done using the API to ensure, for instance, that the exchanged data is encrypted.
\end{itemize}

\begin{figure}
    \centering
    \includegraphics[width=0.48\textwidth]{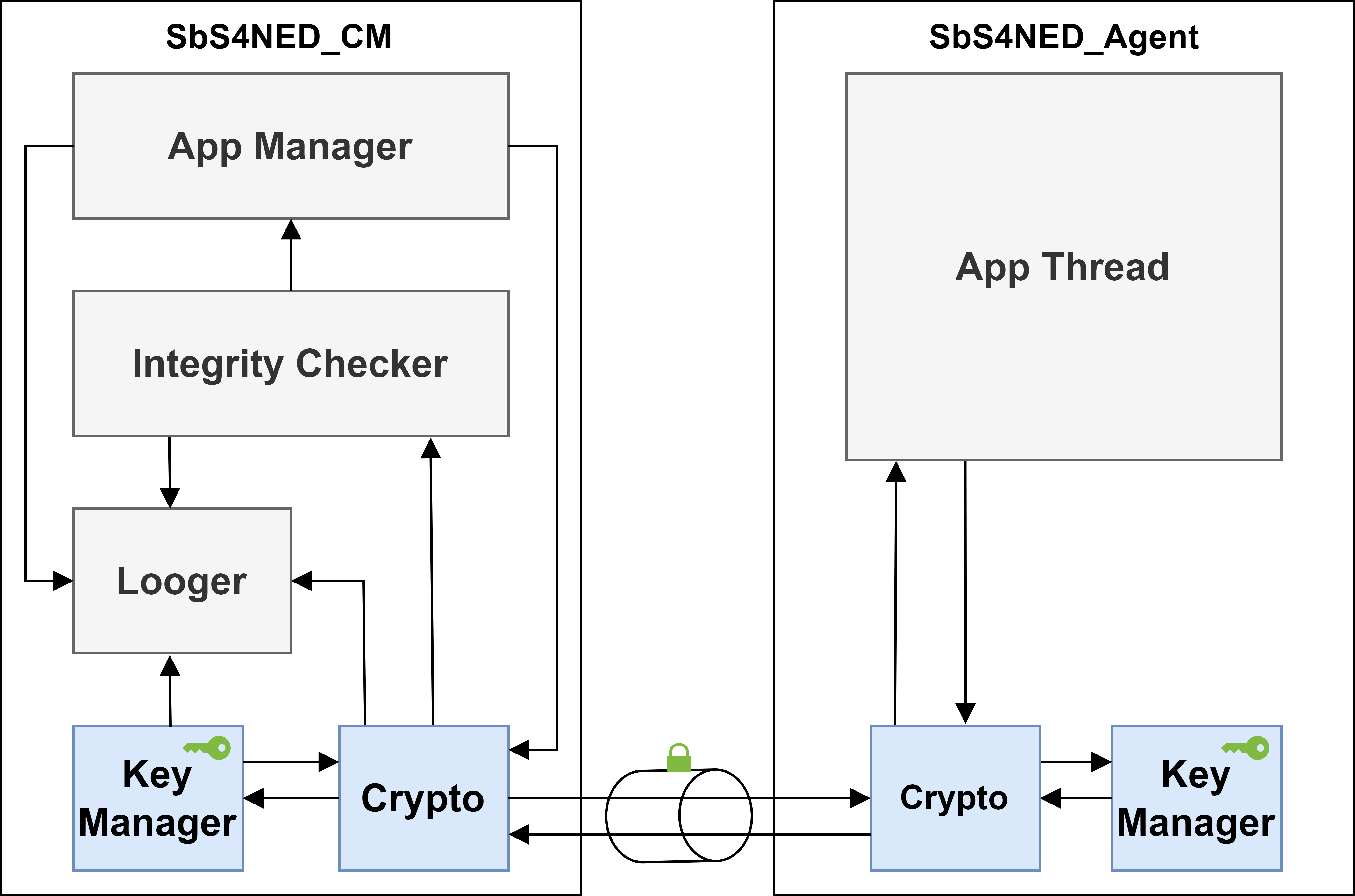}
    \caption{SbS4NED modules.}
    \label{fig:implementationUML}
\end{figure}

The system architecture of SbS4NED\_CM and its agents is illustrated in Figure~\ref{fig:implementationUML}. The figure provides an overview of the individual components and their interconnections within the system.

The Key Manager and Crypto components are the essential modules used in the SbS4NED\_CM and also in its agents. These components are deployed on both sides of the system, providing the necessary encryption services and ensuring that data transmitted between the agents and SbS4NED\_CM remain secure and confidential.

The App Manager, Integrity Checker, and Logger are part of SbS4NED\_CM. These must be deployed in the Gateway. The App Manager plays a crucial role in dealing with the application code deployed in the agents, providing the necessary services to manage, update, and configure them. On the other hand, the Integrity Checker ensures that the code and messages transmitted by the agent are authentic and have not been tampered with.

\subsection{Data Protection}
Memory integrity is nowadays a crucial security concern. The integrity of the data stored in memory is essential to ensure the system's proper functioning and to prevent unauthorized access or manipulation of sensitive information. 
When data is written in the memory, two pieces of information are stored: the value ($v$) that can be accessed by any other external entity, such as an actuator, and the data integrity ($I$) needed to check the integrity of the data. 
Data integrity  $I$ is computed using the MAC, and its purpose is to ensure that the data in memory has not been tampered with or modified. The formula to compute it is $I=$MAC($v \oplus t$), where $t$ can be a timestamp or a random number known only by the integrity checker. The position of the value $I$ in memory is arbitrary, and only the Integrity Checker knows where it is placed.
The choice of MAC over hash functions is because MAC uses a secret key to generate the authentication code. Assuming that SbS4NED\_CM has exclusive access to this private key, only authorized accesses to SbS4NED\_CM can generate the correct MAC result. In contrast, anyone can compute the hash value by identifying the function used. A secret key prevents anyone from computing the hash value and faking the integrity data.
This way, SbS4NED\_CM provides a strong level of security against malicious attacks to the memory and an easy and efficient verification of the memory's  integrity. 

\subsection{Application Protection}
Besides confidentiality protection provided by encryption, the application code is locally stored in the node and decrypted only during execution. Application code also has integrity protection to ensure it remains unmodified even when other parties can access the node. To check the code's integrity, a challenge-response protocol is used, in which the SbS4NED\_CM will send $enc(t)$, an encrypted challenge to the node, where $t$ can be a timestamp or a value randomly generated by the SbS4NED\_CM. The node has to reply with the hash of the whole application code ($ac$) XOR-ed with $t$, i.e., $enc (hash(ac \oplus t))$. Then, the SbS4NED\_CM checks the validity of the result. If the node fails the validation, SbS4NED\_CM could force an update to restore the node application. If the node gives no response, SbS4NED\_CM assumes that the node is lost or compromised and its data is dropped. 

\subsection{Keys Renewal}

A renewal cycle mechanism can extend the system's lifetime. Therefore, before any application update, the key used to encrypt the application code and for communication between the SbS4NED\_CM and node is renewed using a DH protocol. However, the application could take a long time without needing an update. Therefore, the node is provided with an encryption application code, and the SbS4NED\_CM can initialize the DH key exchange with the node to generate the new key even when there is no call of the functionally. The generated key will be used to renew the encrypt of the application code and in further communication with SbS4NED\_CM.

\subsection{Encryption Algorithms}
Since the proposed architecture uses an encryption algorithm and targets low-end devices, the algorithms that can be used must be lightweight, including the cryptographic ones. Thus, this architecture will use NIST lightweight encryption algorithms to protect code and data during message exchange~\cite{nist}. Although by the time of writing this manuscript, the final stage of the NIST competition comprises ten finalists, we are only interested in algorithms that can deal with stream encryption. The main reason is that we want the encrypted code and messages to have the same size as the original ones. Among the ten finalists, few support stream encryption natively. For the SbS4NED architecture, Xoodyak and ISAP schemes with keyed mode association are considered. Since the SbS4NED architecture is modular, other algorithms may also be used.

\section{Proof of concept}

To verify and characterize the proposed architecture, we are currently implementing it. We plan to deploy the architecture in a prototype, enabling system testing in a distributed environment. The prototype will consist of a Gateway on which the SbS4NED\_CM will be running, with connections to multiple nodes where the SbS4NED agents and applications will be deployed.  By conducting tests in a distributed environment, we can ensure that the architecture can effectively handle the communication and data exchange requirements between the Gateway and the nodes. Additionally, we will be able to identify potential issues or limitations during deployment, which will help us refine the architecture further.

\begin{figure}[h]
\centering
   \begin{subfigure}{0.5\textwidth}
   \centering
   \includegraphics[width=0.8\textwidth]{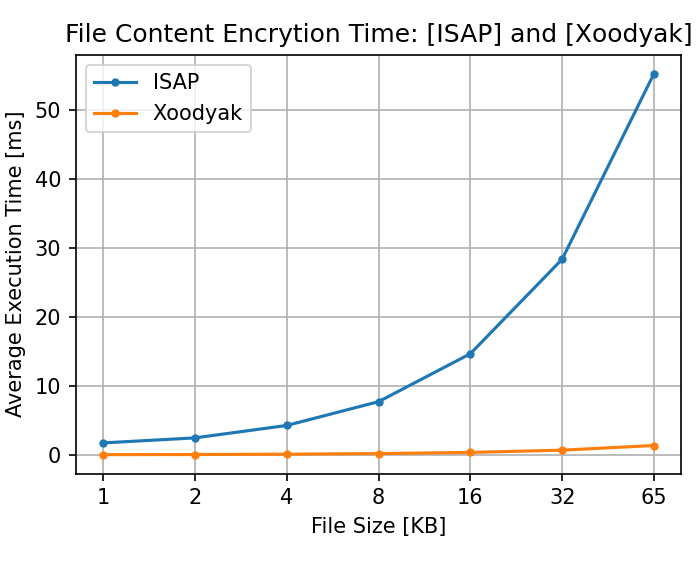}
   \caption{[ISAP] and [Xoodyak]  Encryption on Raspberry Pi 3+.}
   \label{fig:encryption} 
\end{subfigure}
\begin{subfigure}{0.5\textwidth}
\centering
   \includegraphics[width=0.8\textwidth]{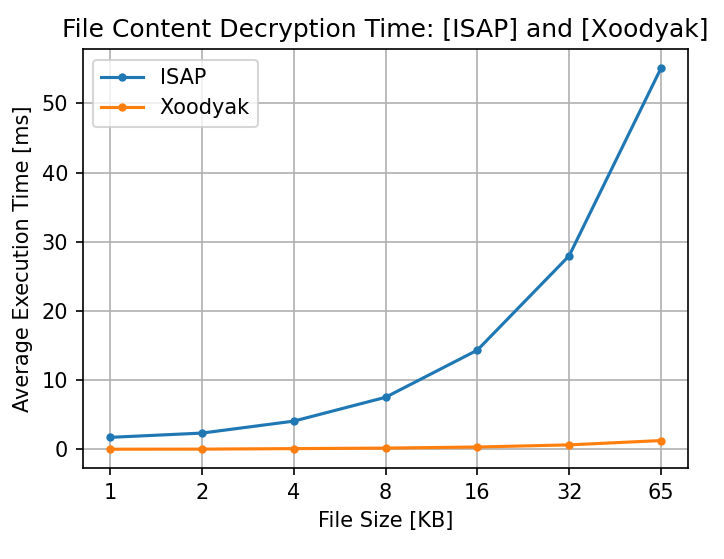}
   \caption{[ISAP] and [Xoodyak] Decryption on Raspberry Pi 3+.}
   \label{fig:decryption}
\end{subfigure}
\caption{[ISAP] and [Xoodyak] on Raspberry Pi 3+.}
\label{fig:encryptionAndDecryptionTime}
\end{figure}

During the early stages of designing the proposed architecture, we conducted experiments using two NIST lightweight encryption algorithms, ISAP and Xoodyak, to determine which would be more suitable for our research. Our experiment used a Raspberry Pi Model 3+ platform with a Quad-core @1.4GHz and 1GB LPDDR2 SRAM. We tested both algorithms for their execution time and memory usage using the same key length of 16 bytes, a nounce of 16 bytes, and file sizes from 1 to 65 kilobytes (kB) (Figure \ref{fig:encryptionAndDecryptionTime}). 
The tests also included both the encryption (Figure\ref{fig:encryption}) and decryption (Figure~\ref{fig:decryption}) processes.

Our experiment shows that Xoodyak was more suitable for our research than ISAP in terms of both average execution time and memory usage. The average execution time for Xoodyak was consistently lower, especially for larger file sizes (generally, Xoodyak is 30 to 60 times faster than ISAP with a maximum standard deviation of 0.004 ms). Moreover, both algorithms required about 370 kb of RAM to perform encryption and decryption tasks. It was noticed that file size does not affect the algorithm's memory usage.
In summary, our experiments with ISAP and Xoodyak algorithms, conducted on a Raspberry Pi Model 3+ platform, helped us to determine which algorithm is more suitable for our research.

\section{Conclusion}
This work proposes a software-based secure execution environment architecture that is lightweight, requires no hardware modifications and is resistant to the most common hardware attacks. Currently, this architecture is being implemented as a proof of concept. After the implementation, tests will be conducted to analyze its performance and characterize its efficiency.

\section*{Acknowledgments}
This work was supported by the LASIGE Research Unit (ref. UIDB/00408/2020 and ref. UIDP/00408/2020), and by the European Union’s Horizon 2020 research and innovation programme under grant agreement No 871259 (ADMORPH project).

\bibliographystyle{ieeetr}

\end{document}